\documentclass[letterpaper]{article}
\usepackage{aaai22}  
\usepackage{times}  
\usepackage{helvet} 
\usepackage{courier}  
\usepackage[hyphens]{url}  
\usepackage{graphicx} 
\usepackage{booktabs}
\usepackage{natbib}  
\usepackage{caption} 
\usepackage{multicol}
\usepackage{subcaption}

\begin{document}

\title{Embedding Elites: Examining the Use of Tweets Embedded in Online News Articles across Reliable and Fringe Outlets\\
\large \textit{MeLa Lab Preliminary Findings Report}}

\author{Benjamin D. Horne\textsuperscript{\rm 1}, Summer Phillips\textsuperscript{\rm 1,2}, and Nelia Koontz\textsuperscript{\rm 1,3}}
\affiliations{\textsuperscript{\rm 1} School of Information Sciences, The University of Tennessee, Knoxville, Knoxville, TN, USA\\
\textsuperscript{\rm 2} Carnegie Mellon University Libraries, Pittsburgh, PA, USA\\
\textsuperscript{\rm 3} University of Wyoming Libraries, Laramie, WY, USA\\
bhorne6@utk.edu, summerp@andrew.cmu.edu, nkoontz@uwyo.edu
}

\maketitle

\begin{abstract}
This study examines the use of embedded tweets in online news media. In particular, we add to the previous literature by exploring embedded tweets across reliable and unreliable news outlets. We use a mixed-method analysis to examine how the function and frequency of embedded tweets change across outlet reliability and news topic. We find that, no matter the outlet reliability, embedded tweets are most often used to relay the opinions of elites, to syndicate information from another news source, or to self-cite information an outlet previously produced. Our results also show some notable differences between reliable media and fringe media's use of tweets. Namely, fringe media embed tweets more and use those tweets as the source of news more than reliable media. Our work adds to the literature on hybrid media systems and the normalization of social media in journalism.
\end{abstract}

\section{Introduction}
Today's media systems are hybrid. As described by Chadwick (2017): professional-driven news cycles develop along side ordinary citizens and activist on social media. It has been argued that this mixing of old and new logics has shifted power away from elites, who in the past retained authority over information flow and could steer the news agenda, to ordinary citizens and grassroots movements. For example, social media and the coverage of information conveyed on social media gave power to anti-government protesters during the Arab Spring \citep{khondker2011role}, and more recently, during the Black Lives Matter movement, activists gained the attention of traditional news media through social media efficacy and coordination \citep{freelon2018quantifying, zhang2018attention}. Hence, hybrid media ecosystems allow for participatory media to be embedded in traditional news media, sometimes fostering a greater civic dialogue than before \citep{lewis2010thinking}. However, it has also been argued that elites, particularly political elites, have regained power in this hybrid setup. For example, hybrid media is frequently used in election campaigns around the world \citep{blach2017hybrid, farhall2019political}, and what an elite individual or institution posts on social media can be inherently ``newsworthy'' simply due to the power they hold as an ``elite'' \cite{harcup2017news}. In this way carefully crafted messaging, and sometimes disinformation, by elites can move across communication contexts and still have agenda setting power.

The role of platforms is central in this space. Most notably, the social media platform Twitter has played a significant role in journalism and news. In fact, one of the first large-scale studies of Twitter debated whether we should consider the platform social media or a type of news media due to the notable number of influential news media accounts on the platform \citep{kwak2010twitter}. In line with this debate, past research has argued that Twitter has ``blurred the lines'' between professional journalists and other types of information producers, ultimately changing journalism practices and norms \citep{olausson2017reinvented,hedman2015j,hedman2013social,parmelee2013political}. 

One such impact, among the many impacts Twitter has had on journalism, is the use of tweets directly in news articles. Better characterizing this practice is the core goal in this paper: \textit{How are tweets used in news articles?} This question is not new, and in fact, a rich set of literature has been centered around the question \citep{broersma2012social, broersma2013twitter, oschatz2021twitter}. However, we argue that this question deserves another look in the new context of fringe, unreliable media. 

Fringe media outlets are amateur outlets branded as alternatives to mainstream media. These outlets are of growing concern to researchers and journalists because they frequently publish low-veracity reports, often based on hyper-partisan conspiracy theories and framing. This fact is not to say that mainstream news organizations have not taken part in the partisan framing of news, but it is to say that fringe media has taken partisan framing to another level. Fringe media outlets have grown audiences across multiple spaces, including social media, online news media, and even cable news media \citep{khudabukhsh2021fringe, munger2022right}, and across those spaces, it has been argued that their content has incited offline harms (i.e. violent events such as the 2016 pizzeria shooting (Pizzagate), the 2021 U.S. Capitol attack, etc.). Although the casual effects of media on ideology and actions are debated and difficult to establish \cite{guess2018avoiding, flaxman2016filter, althoff2017online}, the concerns of fringe media content's effects are warranted. Yet, despite these concerns, researchers have not examined the use of tweets and embedded social media by these outlets. 

To this end, we ask four research questions that replicate previous work in this new context:
\begin{itemize}
    \item \textbf{RQ1:} What are the functions of tweets embedded in news articles?
    \item \textbf{RQ2:} How frequently are tweets embedded in news articles?
    \item \textbf{RQ3:} How do the functions and frequencies of embedded tweets change across news topics?
    \item \textbf{RQ4:} How to research questions RQ1, RQ2, and RQ3 change across news outlet reliability?
\end{itemize}

To explore these questions, we utilize a dataset of online news articles with embedded tweets across reliable and fringe media outlets. We find that news articles that cover topics with clearly defined, public-facing, elite decision makers or figureheads use more tweets than those that do not. When tweets are embedded in news media, they can hold a variety of functions, but most often they are used to relay elite opinions, to syndicate information from another news source, or to self-cite information the outlet previously produced, no matter the outlets veracity. 

While overall there are many similarities in the use of embedded tweets across reliable media and fringe media, there are some notable differences. Proportionally in our data sample, fringe media embeds tweets in their articles approximately eight times more than reliable media and uses those tweets as the source of the article, rather than supporting the article, more often than reliable media. In addition, we find cases in which tweets are used to craft narratives by embedding those tweets out of their original context. 

\section{Related Work}
\subsection{The Role of Twitter in Hybrid Media}~\label{sec:role}
Just as in the work by \citet{oschatz2021twitter} and the work by \citet{dumitrescu2021embedding}, we view this work through the lens of the hybrid media system \citep{chadwick2017hybrid}, which proposes that the dominate media systems of the past have not been replaced by newer media, but rather combined and mixed together. Hence, rather than consumers only getting information from social media platforms, we see legacy media consumed through social media and social media consumed within legacy media, creating an interdependent set of systems. This inter-dependency means it is not a binary choice of where to consume news (social media or professional news media), but a blend of the two, each influencing one another \citep{chadwick2017hybrid}. 

Within this hybrid system, journalist have increasingly relied on Twitter for both communicating information and sourcing it. \citet{heravi2016twitter} found through a survey of journalist in Ireland that both finding leads to stories and sourcing content were prevalent uses of Twitter. \citet{farrar2015controlling} and \citet{bane2019tweeting} note that journalist often ``follow their Twitter feed as though it were the AP news wire, waiting to see what goes viral'',  suggesting that social media can directly influence what events get covered in the news media.


\subsection{The Function of Tweets in News Articles}
Within the more specific topic of tweets embedded in online news articles, several previous works have explored the function of tweets in news articles. Specifically, \citet{broersma2012social, broersma2013twitter} classified tweets in news articles from Dutch and British outlets into four categories: 1. \textit{trigger} - the news articles is about the tweet, 2. \textit{illustration} - the news article contains tweets used as examples for an argument or examples of an opinion, 3. \textit{stand-alone} - ``tweet of the day'' articles where tweets are published on their own, and 4. \textit{Q\&A} - news articles in which tweets are used as question and answer exchanges. \citet{oschatz2021twitter} also use these function categories to analyze tweets in political news articles from German outlets. 

All three of these works demonstrated that tweets are most frequently used as \textit{illustrations} in news articles, followed by use as the \textit{trigger} for an article. While these broad categories have shed light on the function of tweets in news articles, we argue that these categories miss some important nuances in their function. For example, the category \textit{illustrations} encompasses several functions that can be disaggregated. An illustration can be a tweet from a politician (elite opinion) or can be a tweet from a member of the general public (public opinion). Hence, in this work we again ask: \textbf{RQ1:} \textit{What are the functions of tweets embedded in news articles?} However, rather than utilizing these previously studied categories, we take a hypothesis development approach to first find more granular categories, then compare those categories to what was found in the previous literature. 

Additionally, while the functions of embedded tweets have been consistently shown as primarily illustrations in news articles, the frequency of their use has had different results depending on the study and data. As pointed out by \citet{oschatz2021twitter}, these differences occur across countries and different types of media. For example, previous work estimated that more traditional U.S. news outlets used tweets in news coverage between 3\% and 5\% of the time \citep{bane2019tweeting, wallsten2015non, oschatz2021twitter}. When examining web-only U.S. news outlets, this frequency increased significantly to 31\% \citep{bane2019tweeting, oschatz2021twitter}. For the sake of comparing to this wide range of results, we ask: \textbf{RQ2:} \textit{How frequently are tweets embedded in news articles?} Our dataset for analysis is distinct yet comparable to these studies.

Lastly, with the exception of work by \citet{broersma2013twitter}, the majority of works in this area have focused on political news. Hence, we propose another research question: \textbf{RQ3:} \textit{How do the functions and frequencies of embedded tweets change across news topic?} We propose answering this question using a more granular computational approach, rather than the broader qualitative approach used by \citet{broersma2013twitter}. 


\subsection{The Rise of Fringe, Alternative News Outlets}
The definition of what scholars consider alternative media is debated and has changed over time \citep{holt2019key}. In some literature (pre-2015), alternative news has been presented as a positive: democratic, grass-roots, critical media as opposed to hierarchical, corporate, elitism media \citep{fuchs2010alternative}. In the more current literature on disinformation, the definition is quite the opposite, presented as a negative: low-veracity, rumor-spreading, amateur media as opposed to high-veracity, rigorous, professional media \citep{wilson2020cross}. Consistent across these definitions is the absence of professionalism and resources compared to mainstream media. It is also consistent that alternative media has been characterized as ``partisan in its ideology and anti-establishment in its views \citep{chan2017media}.'' In this work, we use the word ``alternative'' to indicate outlets that are of low veracity. We will use the words fringe, alternative, and unreliable interchangeably.

While we and other scholars hesitate to classify some of these fringe outlets as proper journalism, they can occupy the same space that professional, quality journalism does. Recent evidence suggests that consumption has shifted from mainstream to fringe outlets, particularly on the U.S. political right. Specifically, \citet{khudabukhsh2021fringe} found that viewership of Newsmax, a far-right alternative news source, ``increased over a factor of seven'' before the January 6th U.S. capitol riots. There are also practical examples of political elites propping up the perceived legitimacy of fringe media outlets. For example, in 2017, then U.S. President Donald Trump gave the far-right outlet One America News Network a permanent seat in the White House briefing room\footnote{https://tinyurl.com/3kt6hpy4}. 


Therefore, understanding fringe news media's practices is just as important as, if not more so than, understanding quality news media's practices. This reasoning motivates our last and primary research question, \textbf{RQ4:}\textit{ How do research questions} \textbf{RQ1}, \textbf{RQ2}, \textit{and} \textbf{RQ3} \textit{change across news outlet veracity?} With regard to the work on tweets embedded in news articles, studies consistently found that quality newspapers used tweets more often than tabloid newspapers \citep{oschatz2021twitter}. However, this set of findings was in the context of political coverage and over a much narrower range of quality than what we examine in this work. 


\begin{table}[ht!]
\centering
\fontsize{9.2pt}{9.2pt}
\selectfont
\begin{tabular}{c|c|c}
\toprule
\multicolumn{3}{c}{\textbf{Reliable}}\\\toprule
   \textbf{Outlet} & \textbf{\# Articles} & \textbf{\% w/Tweets}\\\midrule
    New York Times & 13666 & 0.70\% \\
    NPR & 3167 & 0.079\% \\
    ABC News & 3867 & 0.00\% \\
    USA Today & 8288 & 0.24\% \\
    The Hill & 2020 & 22.42\% \\
    The Guardian & 1543 & 4.93\%\\
    CNN & 17737 & 1.81\% \\
    CBS News & 26253 & 3.1\% \\
    Washington Post & 2483 & 0.36\%\\
    \textbf{Total Reliable} & \textbf{80629} & \textbf{2.26\%}\\
    \midrule
    \multicolumn{3}{c}{\textbf{Fringe}}\\\toprule
   \textbf{Outlet} & \textbf{\# Articles} & \textbf{\% w/Tweets}\\\midrule
    Activist Post & 1478 & 13.32\%\\
     Veterans Today & 1632 & 20.03\%\\
     Gateway Pundit & 4556 & 44.89\%\\
     Newsmax & 3391 & 0.00\%\\
     OAN & 1647 & 27.69\%\\
     Breitbart & 9450 & 31.53\%\\
     True Pundit & 4495 & 26.92\%\\
     Shareblue & 1154 & 0.001\%\\
     News Busters & 13053 & 7.02\%\\
   \textbf{Total Fringe} & \textbf{40856} & \textbf{19.90\%}\\
\end{tabular}
    \caption{Metadata on each source in our sampled dataset. Note, 84\% of the tweets embedded by reliable outlets were from Twitter verified accounts, while 75\% of the tweets embedded by fringe outlets were from verified accounts.}
    \label{tab:metadata}
\end{table}

\section{Data and Methods}\label{sec:methods}
\subsection{Data Sample}\label{sec:data}
To answer our research questions, we extract subset of data from the NELA-GT-2021 dataset containing every news article published from 18 outlets between January 2021 and May 2021 \citep{gruppi2022nela}. The process we used to select this subset is as follows:
\begin{enumerate}
    \item We randomly select nine ``reliable'' outlets and nine ``unreliable'' outlets from the dataset. To do this, we utilize outlet reliability labels aggregated in the dataset. Gruppi et al. (2022) assign news outlets one of three labels: \textit{unreliable}, \textit{mixed}, and \textit{reliable}. These labels are based on the ``factuality ratings'' provided by Media Bias/Fact Check\footnote{https://mediabiasfactcheck.com/}. Using the 348 outlets labeled in the dataset, we randomly select nine from the reliable/mixed group (called \textit{reliable} henceforth) and nine from the unreliable group (called \textit{fringe} henceforth). We choose to randomly select these outlets, rather than hand-select them, to avoid any potential bias on our part in the process.
    \item With these selected outlets, we extract the full text from all articles, including any embedded tweets, between January and May. Given the completeness of the NELA-GT-2021 dataset, this subset should include nearly every article published during this time-frame.
\end{enumerate}



Following this process, we extracted a total of \textbf{121,485 articles} from \textbf{18 news outlets} over \textbf{five months}. Of those 121,485 articles, 9,952 (8.2\%) of them contained at least one embedded tweet. This data broken down by outlet can be found in Table~\ref{tab:metadata}.

\subsection{Structured Topic Modeling}\label{sec:stm}
Our analysis methodology is made up of two major parts, the first being topic modeling. Namely, we run a Structured Topic Model (STM) \cite{roberts2014structural} on our full dataset, with one binary covariate in the model, representing if the article had an embedded tweet or not. A STM is a generative model of word counts that allows for document-level metadata to be used in the model's creation. As with most topic models, this process produces a topic-word distribution per document, however, with a STM we can associate each topic-word distribution with other variables. This additional piece of information allows us to examine how a variable effects the themes in each document. Hence, this method is ideal for our research questions as we can associate topics within the news as more or less likely to contain an embedded tweet at scale. We choose to use a STM over other topic models not only because of its clear applicability to our research questions, but also because the method can be easily replicated \citep{rodriguez2020computational}.

In many topic models, including STMs, the number of topics extracted is a tunable parameter (often called $k$). To select $k$ we use a standard method used across many topic modeling studies. That is, we use a grid search over values of $k$ to find the topic model that best predicts topic probabilities on a held-out sample of words, measured by the minimum perplexity. This process indicates that the best fitting model on our full data set has 73 topics. In addition, we use standard set of preprocessing steps on our text. These preprocessing steps create a set of normalized, content relevant words per article. For a brief overview of text preprocessing, please see \citet{kannan2014preprocessing}. For an introduction to topic modeling, please see \citet{blei2003latent}.

\begin{figure}[ht!]
    \centering
    {\includegraphics[width=6.5cm]{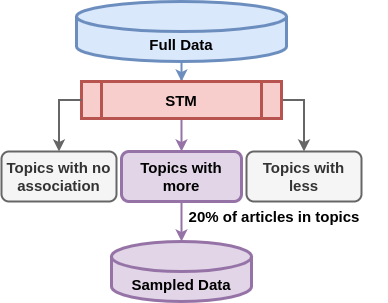}}%
    \caption{Steps in our qualitative sampling method, where articles are sampled from the 12 topics that were associated with more embedded tweets.}%
    \label{fig:flow}%
\end{figure}

\subsection{Open Coding of the Functions of Tweets}\label{sec:qual} 
The second major part of our analysis methodology is manual, rather than automated. We employ an inductive coding process to categorize the function of tweets in articles. To do this, we create a sample of our full dataset that is more suitable for qualitative work.

\paragraph{Sample}
Rather than randomly sampling over the full dataset, we choose to utilize the topic model described above in a unique way. Specifically, we use the model not only to evaluate which topics are more or less associated with the use of embedded tweets, but also as a step in our sampling process. We sample articles from each topic that is associated with having significantly more embedded tweets according to the model. To do this, we collapse the topic vector per document into a single document-to-topic mapping by only keeping the topic with the largest proportion associated with the document (we call this the \textit{dominate topic}). From this one-to-one mapping, we filter down the dataset to only include articles whose dominate topic is significantly associated with the use of embedded tweets (articles in 12 of 73 topics). For each of these 12 topics, we sample 20\% of the articles mapped to the topic. A flow chart of this process is shown in Figure \ref{fig:flow}.

By creating our qualitative coding dataset using the topic groups rather than randomly over the full dataset, we can sample systematic variation in content and perform our qualitative analysis across the topics of interest, even if those topics are only represented by a relatively small number of articles. This choice also allows us to map the output of our inductive coding back to the topics produced by the STM.

In the end, this sampling process gave us a dataset of 410 articles across the 12 of the 73 topics. Note, because the sample is done per topic document group, and only those topics associated with higher number of tweets, our sample size is much less than a 20\% sample size of the data set as a whole. 

\paragraph{Coding}
We then coded these embedded tweets using a two-step coding process. First, two authors individually went over the sampled articles to identify broad codes of how the tweet(s) were being used in each article. The two coders and one other author then met to discuss the coding and develop a combined list of codes. These identified codes can be found in Table ~\ref{tab:codes}. Next, the two coders went back over all of the sampled data to re-code them based on the agreed upon tweet function categories from step one. 

Importantly, tweets can be assigned to more than one category and articles can contain more than one tweet. Furthermore, as shown in our developed codes in Table~\ref{tab:codes}, every article is assigned at least one of eight specific function codes (i.e. \textit{What are the tweet(s) in the article?}) and at least one of two codes representing if the embedded tweet(s) were the \textit{source} of the article or \textit{supporting} the article (i.e. \textit{How do the tweet(s) relate to the article?}). The `source' and `support' codes are similar to the \textit{trigger} and \textit{illustration} categories developed by \citet{broersma2012social, broersma2013twitter}.

After the second round of coding, the agreement between the two coders across all sampled articles for the first group of eight codes (What are the tweet(s) in the article?) was 83.00\% and for the second group of two codes (How do the tweet(s) relate to the article?) was 99.02\%. An article was said to be agreed on if all assigned codes for the article aligned between the coders. Disagreements were settled to reach 100\% agreement for analysis. 




\begin{table*}[ht!]
\centering
\fontsize{9.2pt}{9.2pt}
\selectfont
\begin{tabular}{c|p{11.8cm}}
\toprule
\multicolumn{2}{c}{\textbf{What are the tweet(s) in the article?}}\\\midrule
 \textbf{Code} & \textbf{Code Description} \\\midrule
   Official Statement &  Tweet from an organization or public official making a statement on an issue. Examples of organizations include political organizations and sports leagues. \\\midrule
   Elite Opinion &  Tweet from verified, highly followed Twitter users. Examples include politicians and celebrities.\\\midrule
   Public Opinion &  Tweet from unverified Twitter users, framed as public opinion in the article.\\\midrule
   Journalist/Citizen Journalist & Tweet from professional journalist or amateur journalist. Importantly, amateur journalist is used broadly here: if a twitter user claims to be a journalist in their profile the tweet is placed in this category. \\\midrule
   Tweet Syndication &  Tweet from a different news outlet than the article itself. Note, the tweet is from a news outlets Twitter account, rather than a journalists Twitter account. \\\midrule 
   Internal Link &  Tweet from the Twitter account of the same news outlet that the article is from. These are often pointing to related articles or video/cable coverage of the news in the article.\\\midrule
   Live Event & Tweet documenting a live event, typically with video. These can be citizens or journalist on the ground.\\\midrule
   Curated List & The article contains a list of embedded tweets on a specific subject. Most often these lists are of elite opinions, however, in several cases these lists are of tweets containing live videos. An example can be found in Figure~\ref{fig:example}.\\\bottomrule\toprule
   \multicolumn{2}{c}{\textbf{How do the tweet(s) relate to the article?}}\\\midrule
   \textbf{Code} & \textbf{Code Description} \\\midrule
   Source (\textit{trigger}) &  The news article is about the tweet(s) or the tweet(s) are used as the source of the article\\\midrule
   Support (\textit{illustration}) & The news article uses the tweet(s) to support or illustrate a point made in the article\\\bottomrule
\end{tabular}
\caption{Codes developed through qualitative process described in Section \ref{sec:qual}. Note, every article is labeled with at least 2 codes: at least one from Source or Support, and at least one from the remaining codes. }
\label{tab:codes}
\end{table*}

\begin{table}[ht!]
\centering
\fontsize{8.2pt}{9.5pt}
\selectfont
\begin{tabular}{c|c|c|c}
\toprule
  \textbf{Category} & \textbf{Reliable Outlet} & \textbf{Fringe Outlet} & \textbf{Sig}\\\midrule
    Official Statement & \textbf{30.6\%} & 20.6\% & ***\\
    Elite Opinion & 44.9\% & \textbf{55.6}\% & ***\\
    Public Opinion & 11.2\% & 13.5\% & \\
    Journalist/Citizen Journal & 17.4\% & \textbf{29.3\%} & ***\\
    Tweet Syndication & 21.4\% & \textbf{34.4\%} & ***\\
    Internal Link & \textbf{7.1\%} & 2.9\% & ***\\
    Live Event & 21.4\% & 16.1\% & \\
    Curated List & 7.1\% & \textbf{15.4\%} & ***\\\midrule
    Source & 52.0\% & \textbf{77.8}\% & ***\\
    Support & \textbf{66.3}\% & 52.4\% & ***\\\bottomrule

\addlinespace[1ex]
\multicolumn{3}{l}{\textsuperscript{***}$p<0.01$, 
  \textsuperscript{**}$p<0.05$}
  \end{tabular}
\caption{Percentage of each coded Tweet use category for each outlet type. The column `Sig' shows if the observed imbalance between reliable and fringe outlets within a category is statistically significant according to a Fisher Exact test. Percentages are in bold if they are significantly greater than percentage in the opposing outlet type. Final coding is done on a sample balanced across topics, totalling in 410 articles across the 12 topics where tweets were used significantly more. Note, an article's use of a tweet can be placed in more than one category or an article can have more than one embedded tweet.} 
\label{tab:codes_outlettype}
\end{table}

\section{Results}

\subsection{The frequency and function of embedded tweets} \label{sec:freq_and_func} 
\paragraph{Frequency} In Table \ref{tab:metadata}, we show the percent of articles with embedded tweets per news outlet and outlet type. In this Table, two clear findings stand out. First, fringe outlets embed proportionally more tweets in their news articles than reliable outlets. Namely, only 2.26\% of articles from reliable outlets contained at least one embedded tweet (1822 of 80629 articles), while 19.90\% of articles from fringe outlets contained at least one tweet (8130 of 40856 articles). 

Second, the frequency of tweets used in articles is not uniform across outlets in each group. For example, in the reliable group, The Hill uses tweets in their news articles much more frequently than the rest of the outlets in the group, with 22.42\% of articles having at least one tweet. The next outlet that most frequently uses tweets is The Guardian, with only 4.93\% of its articles containing a tweet. In fact, five of the nine sources in the reliable group use tweets in less than 1\% of their articles, significantly different frequency than The Hill. In the fringe group, the use of tweets is more balanced across outlets, but a few stand out as using tweets much less than the rest. Newsmax and Shareblue both use embedded tweets almost never (0.00\% and 0.001\% respectively), while six of the nine fringe outlets use tweets in greater than 13\% of their articles. 


\paragraph{Function} In Table \ref{tab:codes}, we show the function codes developed through our inductive coding process described in the Methods Section. In Table \ref{tab:codes_outlettype}, we show how these codes change across outlet type. There are several key findings from these tables (Tables \ref{tab:codes} and \ref{tab:codes_outlettype}). First, broadly speaking, tweets embedded in news articles were used to relay opinions or were used to point to other outlet's news, either by embedding other news organization's tweets or other journalist's tweets. Second, while all functions are used by both types of outlets, there are significant differences in how often they are used. Reliable outlets used embedded tweets as official statements from organizations and as internal links to their previously published news significantly more often than fringe outlets. On the other hand, fringe outlets used embedded tweets to relay opinions and syndicate news from others significantly more than reliable outlets. 

\begin{figure*}[ht!]
    \centering
    \subfloat[\centering Article header]{{\includegraphics[width=6cm]{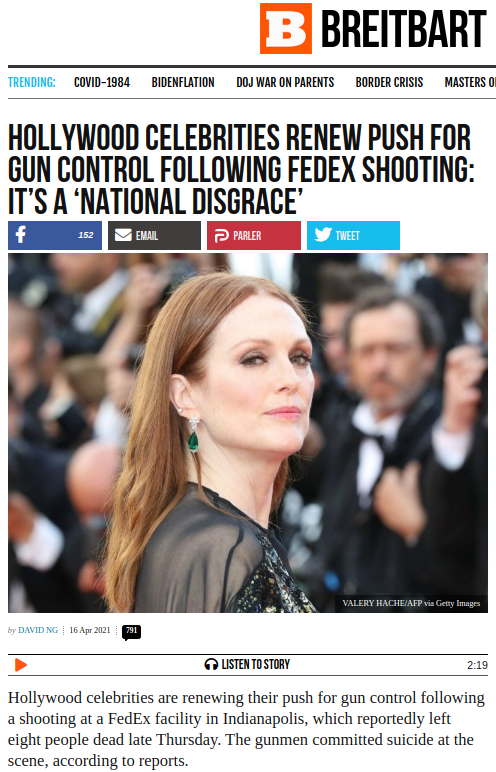} }}%
    \qquad
    \subfloat[\centering Embedded tweets in article]{{\includegraphics[width=5.8cm]{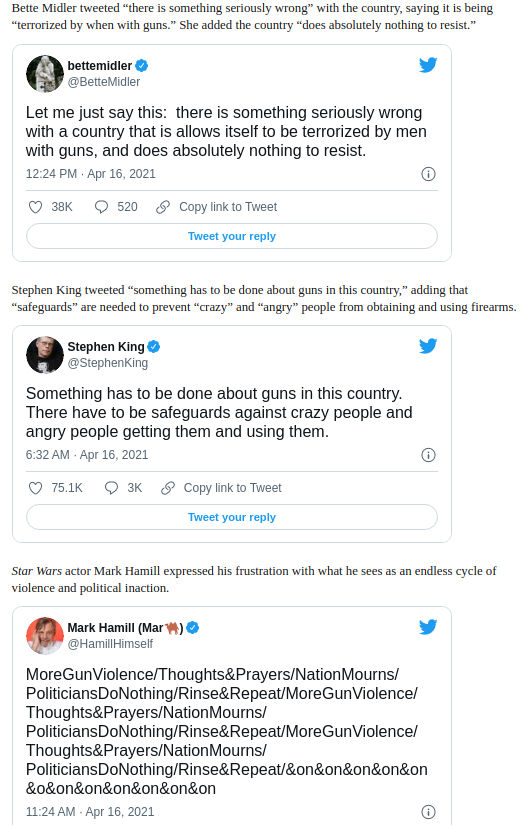} }}%
    \caption{Example of `Curated List' category described in Table~\ref{tab:codes}. In this example, we have an article from Breitbart that contains a curated list of Hollywood celebrities opinions expressed on Twitter. While we show three in this image, the article contains eight embedded tweets of this nature.}%
    \label{fig:example}%
\end{figure*}

Despite these differences, across both reliable and fringe outlets, embedded tweets were most often used to relay elite opinions, although fringe outlets embedded elite's opinions significantly more than reliable outlets. We saw elite opinions covered in two forms throughout our analysis. Most often elite opinions are covered with one or two embedded tweets which serve as the source of an article or support the main story of the article. Less often, and more common in fringe outlets, is what we called `curated lists'. Articles that fell into the `curated list' category were articles that contained very little text (sometimes less than three sentences) and many embedded tweets (between five and 15 tweets). An example of an article in this category can be found in Figure \ref{fig:example}. These tweets were often organized as if the tweets were a conversation between elites on a specific topic, even if they were not directly interacting with each other on Twitter. Sometimes these list were all relaying the same opinion and others were differing opinions framed as `Twitter battles'. In fringe outlets, the little text that is in the article was often the opinion of the writer, which could be disparaging if the writer disagreed with the list of opinions. In reliable outlets, the articles were typically more straightforward, simply stating the consensus opinion among a group. 

These results are further supported when examining what accounts were most frequently embedded across the full dataset (9952 articles, before filtering by the topic model). That is, the top 25 most frequently embedded accounts across both outlet types contain many elites (shown in Figure \ref{fig:topaccounts}). In reliable articles, politicians such as Ocasio-Cortez, Cruz, Trump, Greene, and Biden make the top 25. We also see elites who are not politicians appear in the top 25, such as Elon Musk. In fringe articles, we saw a similar set of politicians in the top 25: Cruz, Trump, Ocasio-Cortez, Greene, and Abbott. Despite the high overlap between the users embedded in these different article types, the context and content surrounding those tweets can be different. For example, both fringe and reliable outlets embedded tweets by Greene when covering her removal from several congressional committees, but the writing and framing around those tweets were different across outlet types. 

Mixed in with the elites in the top 25 most frequently embedded accounts were news accounts, such as CBSThisMorning, FaceTheNation, and CNN for reliable articles, and BreitbartNews, DailyCaller, thehill, and CBSNews for fringe articles. Note that fringe outlets were often embedding tweets from reliable outlets in their articles, aligning with the `Tweet Syndication' category found in our inductive coding process (used in 21.4\% of articles by reliable and 34.4\% of articles by fringe, see Table~\ref{tab:codes_outlettype}). Additionally, the frequency of news accounts embedded aligns with the `Internal Link' category found in our coding process (used in 7.1\% of articles by reliable and 2.9\% of articles by fringe).

Lastly, we found that the use of tweets as the \textit{source} of the article or to \textit{support} the article differs across outlet type. As shown in Table \ref{tab:codes_outlettype}, reliable outlets more often used tweets to support the article, while fringe outlets used the tweets more often as the source of the article. However, there are many cases across both outlet groups where one tweet is used as the source and another tweet is used as support.


\subsection{How do the functions and frequencies of embedded tweets change across news topic?}\label{sec:topic_change} 

As expected, both the function and the frequency of tweets change across topics. In Table \ref{tab:stm} we show the extracted topics from the STM that were significantly associated with more or less embedded tweets compared to other topics extracted from the full set of data. As a reminder, significance is determined by the STM, which indicates when a metadata variable is significantly associated with an extracted topic. For more information on Structured Topic Models, please see the R package documentation\footnote{\url{https://cran.r-project.org/web/packages/stm/vignettes/stmVignette.pdf}}. For the sake of space, if a topic had no relationship with the number of embedded tweets, we did not include it in the table. 


Topics associated with more tweets than average fell into one of three groups: political news, big business news, and celebrity news. More precisely, topics on policy making and decisions (i.e. COVID-19 mitigation, state-level policies, congress voting outcomes, Trump's impeachment) and topics on specific political events/scandals (i.e. January 6th Capitol riot, Cuomo sexual assault allegations) used more tweets. News stories about big business products and policies (i.e. automotive industry, airline industry), sporting events, or celebrities also used more tweets.

In Table \ref{tab:topic_to_code}, we show the percentage of articles in each topic that fall into each of the codes developed in our qualitative analysis. We only show this breakdown for topics that were associated with significantly more tweets. From this Table, as expected from our previous results, we can see that `Elite Opinion' was most often used across topics. 

Broadly, these results suggest that topics which cover situations with clearly defined, public-facing, elite decision makers or figureheads use significantly more tweets. For example, while many political topics use embedded tweets, political topics that had no clear decision maker or figurehead, such as the removal of confederate monuments (topic number 7 in Table \ref{tab:stm}) use significantly less tweets than the average. Similarly, news covering social movements, such as the Black Lives Matter movement (topic number 70 in Table \ref{tab:stm}), where there was not a clearly defined, central figurehead tweeting about the movement, use significantly less tweets. There are some clear exceptions to this generalization, such as news articles that fall into our 'Curated List' category, where many different elite and public opinions are embedded into articles, or our 'Journalist/Citizen Journalist' category where the embedded tweets are from journalist (or self-proclaimed journalist) covering an event.

\begin{table*}[ht!]
\centering
\fontsize{9.2pt}{9.2pt}
\selectfont
\begin{tabular}{c|c|c|c} 
\toprule
    \textbf{Topic \#} & \textbf{Stemmed Topic Words} & \textbf{Tweets} & \textbf{Topic Interpretation} \\\midrule
    3 & capitol, riot, offic, attack, jan, violenc, support & +*** & Capitol riot\\
     27 &  car, vehicl, drive, driver, crash, electr, truck & +*** & Automotive industry, car crashes  \\
    37 & mask, wear, vaccin, cdc, peopl, fauci, covid  & +*** &  COVID-19 mitigation\\
    42 & citi, new, york, florida, resid, mayor, restaur  & +*** & State-level policies\\
    46 & senat, vote, democrat, impeach, republican, trial & +* & Trump impeachment\\
     49 & famili, year, old, mother, life, children, friend & +*** & Mixed family-related news, Rush Limbaugh's death\\
     54 & cuomo, governor, york, new, alleg, sexual, investig  & +** & Cuomo assault allegations\\
    55 & film, actor, movi, music, star, award, hollywood  & +** & Celebrities, movie, music\\
    58 & gun, shoot, asian, violenc, crime, mass, weapon & +. & Sinophobic crimes\\
    65 & hous, republican, member, cheney, democrat, vote & +*** & Congress votes\\
     72 & travel, flight, airlin, plane, airport, air, fli & +* & Travel, Covid-19 related travel\\
    73 & fight, paul, battl, mike, minut, beat, fought & +* & Boxing/MMA/UFC\\\midrule
    7 & american, histori, war, nation, year, countri, presid & -*** & Confederate Monuments\\
    11 & compani, busi, million, work, worker, fund, organ  & -** & Employment\\
    22 & russia, state, iran, russian, foreign, nuclear, unit & -* & Mixed foreign news\\
    41 & school, student, educ, teacher, univers, children & -. &  Education\\
    52 & court, justic, case, attorney, law, legal, rule & -** & Lawsuits\\
    61 & covid, pandem, health, coronavirus, death, case  & -** & Covid-19 deaths \& care\\
    70 & black, white, peopl, racial, communiti, race, racism  & -** & Racism, social movements, BLM movement\\
    \bottomrule
\addlinespace[1ex]
\multicolumn{3}{l}{\textsuperscript{***}$p<0.01$, 
  \textsuperscript{**}$p<0.05$, 
  \textsuperscript{*}$p<0.1$}
\end{tabular}
\caption{Topics with significantly more or less embedded tweets. In the column Tweets, we show a $+$ for topics with more tweets or a $-$ for topics with less tweets. Additionally, we show $*$ to indicate the p-value of the relationship according to the STM. If a topic showed no relationship with embedded tweets, it is not included. The column Topic Interpretation is decided on by two authors examining both the topic words and articles mapped to the topic.} 
\label{tab:stm}
\end{table*}

\begin{table*}[ht!]
\centering
\fontsize{9.2pt}{9.2pt}
\selectfont
\begin{tabular}{c|c|c|c|c|c|c|}
\toprule
& \multicolumn{6}{c}{\textbf{Topic numbers}}\\\midrule
   \textbf{Category} & \textbf{3} & \textbf{27} & \textbf{37} & \textbf{42} & \textbf{46} & \textbf{49} \\\midrule
    Official Statement & 19.4\% & 13.3\%& 18.9\% & 25.0\% & 24.3\%& 21.6\%\\
    Elite Opinion & \textbf{50.0\%}& 40.0\% & \textbf{67.6}\%& \textbf{30.6}\%& \textbf{56.8\%}& \textbf{56.8\%}\\
    Public Opinion & 8.3\%& 23.3\%& 8.1\%& 5.6\%& 16.2\%& 13.5\%\\
    Journalist/Citizen Journalist & 33.3\% & 10.0\%& 21.6\%& 16.6\%& 43.2\%& 27.0\%\\
    Tweet Syndication & 22.2\%& 10.0\%& 37.8\% & 27.5\% & 21.6\%& 18.9\%\\
    Internal Link & 0.0\%& 3.3\%& 5.4\%& 2.8\%& 0.0\%& 2.7\%\\
    Live Event & 30.6\%& \textbf{50.0\%}& 2.7\%& \textbf{30.6\%}& 8.1\%& 13.5\%\\
    Curated List & 19.4\%& 0.0\%& 13.5\% &13.9\% & 18.9\%& 13.5\%\\\midrule
    Source & \textbf{75.0\%}& \textbf{60.0\%}& \textbf{67.6\%}& \textbf{69.4\%}& \textbf{83.8\%}& \textbf{67.6\%}\\
    Support & 61.1\%& 43.3\%& 51.4\%& 44.4\%& 54.1\%& 51.4\%\\\bottomrule
    \toprule
    & \multicolumn{6}{c}{\textbf{Topic numbers}}\\\midrule
   \textbf{Category} & \textbf{54} & \textbf{55} & \textbf{58} & \textbf{65} & \textbf{72} & \textbf{73} \\\midrule
    Official Statement & 29.7\% & 24.3\% & 27.0\%& 13.5\%& 29.7\%& 20.0\%\\
    Elite Opinion & \textbf{59.5}\%& \textbf{73.0\%}&\textbf{46.0\%} & \textbf{73.0\%}& 29.7\%& 33.3\%\\
    Public Opinion & 16.2\%& 18.9\%&24.3\% & 0.0\%& 10.8\%& 33.3\%\\
    Journalist/Citizen Journalist & 43.2\%& 5.4\% & 13.5\%& 46.0\%& 27.0\%& 20.0\%\\
    Tweet Syndication & 51.4\%& 24.3\%& 27.0\%& 51.4\%& 29.7\%& \textbf{60.0}\%\\
    Internal Link & 8.1\%& 0.0\%& 16.22\%& 2.7\%& 2.7\%& 0.0\%\\
    Live Event & 0.0\%& 0.0\% & 5.4\%& 0.0\%& \textbf{32.4}\%& \textbf{60.0\%}\\
    Curated List & 8.1\% & 27.0\% & 13.5\%& 10.8\%& 8.1\%& 0.0\%\\\midrule
    Source & \textbf{75.7\%}& \textbf{83.8\%}& \textbf{67.5\%} &\textbf{86.5\%} & 48.6\%& 40.0\%\\
    Support & 73.0\%& 45.9\%& 51.3\% & 40.5\%& \textbf{86.5\%} & \textbf{66.7\%}\\\bottomrule
\end{tabular}
\caption{Percentage of each coded tweet function category for each topic. The largest category represented in each topic is in bold. The coding is done on a sample balanced across topics, totalling in 410 articles across the 12 topics where tweets were used significantly more. Note, an article's use of a tweet can be placed in more than one category and/or an article can have more than one embedded tweet. Each topic can include articles from both reliable and fringe outlets.} 
\label{tab:topic_to_code}
\end{table*}




\subsection{Who are the elites embedded in news articles?}
Given that our results have indicated that tweets by elite actors were most often embedded in news articles, we wondered who these elites were. To take a deeper dive, we performed a second inductive coding analysis to categorize what type of elites are embedded. This process followed the same multi-round approach as our first coding process, however, the coding was only done on articles previously coded as containing `Elite Opinion' in our first coding analysis (115 articles). Hence accounts such as news accounts or journalist accounts are not included in the analysis. After two rounds of coding, the agreement between the two coders was 100\%. In Table \ref{tab:elites_coding}, we display the percent of articles that contain one of six broadly identified account types from our second round of qualitative coding. 
\begin{table}[ht!]
\centering
\fontsize{9.2pt}{9.2pt}
\selectfont
\begin{tabular}{c|c}
\toprule
  \textbf{Elite Category} & \textbf{\% of Articles} \\\midrule
    Politicians & \textbf{72.8\%}\\
    Hollywood Celebrities  & 19.3\%\\
    Athletes & 2.6\%\\
    Activist & 2.6\%\\
    Billionaires & 1.8\% \\
    Book Authors & 0.9\%
\end{tabular}
\caption{Coded categories of elites from articles coded as having tweets relaying elite opinion.}
\label{tab:elites_coding}
\end{table}

Aligning with multiple of our previous results, we found that the vast majority of elites embedded in news articles are politicians (72.8\%), followed by Hollywood celebrities (19.3\%). Much less frequently, we found athletes, famous activists, and billionaires embedded in articles.

\begin{figure*}[ht!]
    \centering
    \subfloat[\centering Reliable outlets]{{\includegraphics[width=5.3cm]{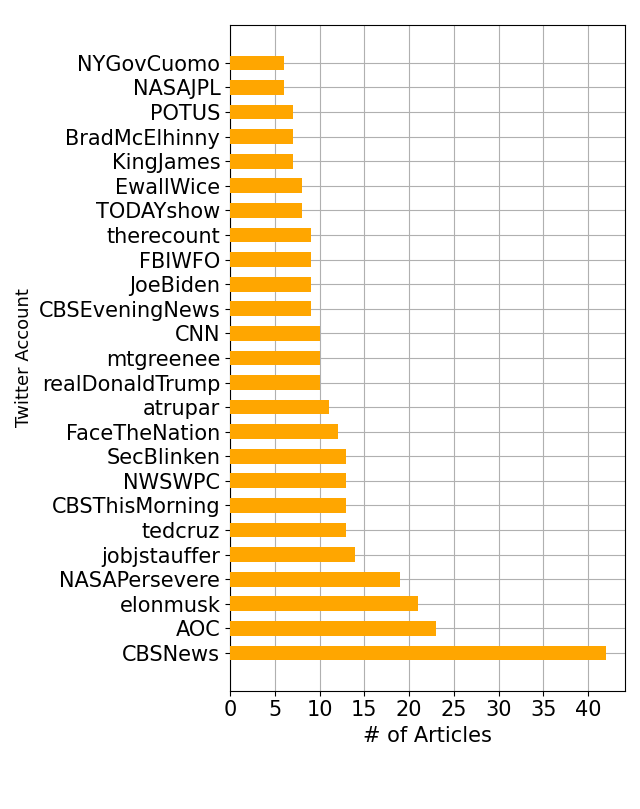} }}%
    \qquad
    \subfloat[\centering Fringe outlets]{{\includegraphics[width=5.3cm]{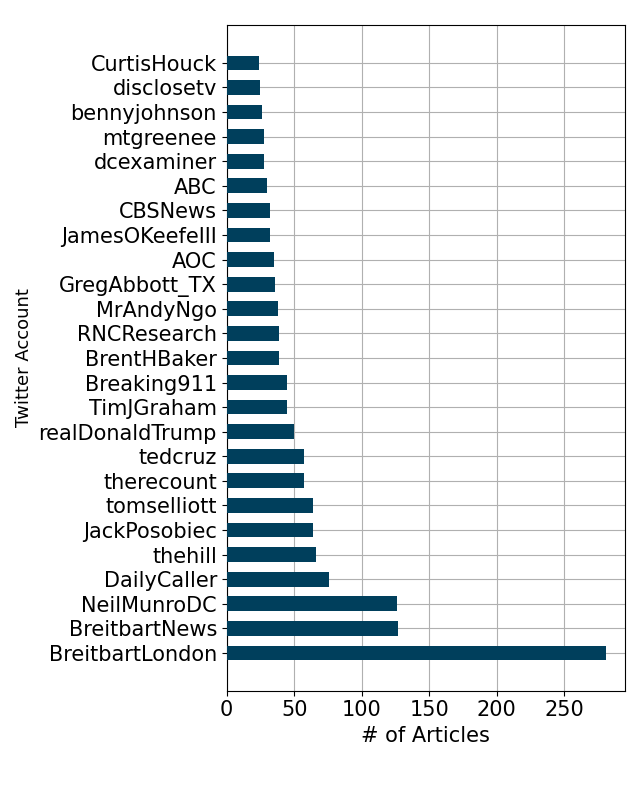} }}%
    \qquad
    \subfloat[\centering All outlets]{{\includegraphics[width=5.3cm]{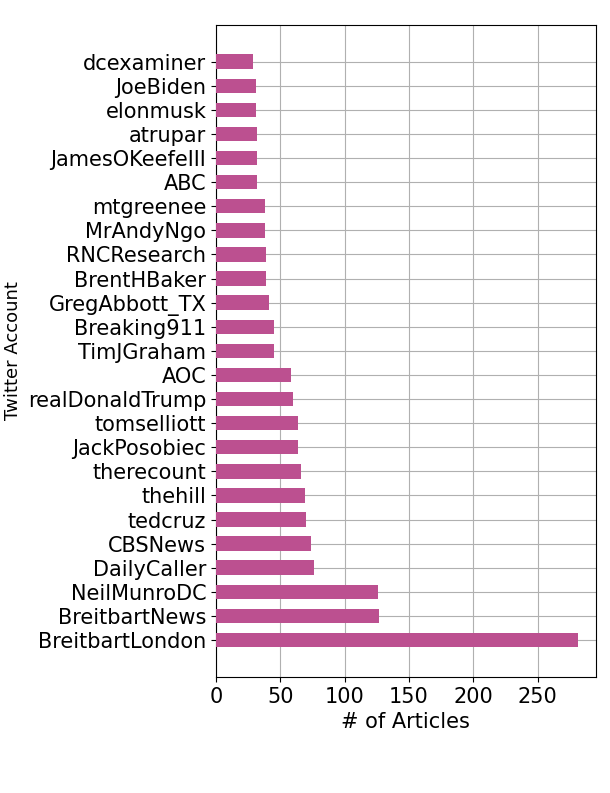} }}%
    \caption{Top 25 embedded Twitter accounts for (a) Reliable (b) Fringe outlets, and (c) all outlets. This count was done across the full dataset of 9952 articles with embedded tweets. Importantly, the frequency of an account being embedded does not perfectly align with the number of articles that one is embedded in, as multiple tweets can be embedded in one article. Further, note that the X-axis are at different scales.}%
    \label{fig:topaccounts}%
\end{figure*}

\begin{table*}[ht!]
\centering
\fontsize{8.0pt}{8.5pt}
\selectfont
\begin{tabular}{c|c|c} 
\toprule
    \textbf{\#} & \textbf{Topic Interpretation} &\textbf{Top 5 Twitter Accounts} \\\midrule
    3 & Capitol riot & realDonaldTrump, Julio\_Rosas11, BradMcElhinny, atrupar, tedcruz\\
     27 &  Automotive, crashes & elonmusk, willhaynie, DJJudd, NASCAR, TigerWoods\\
    37 &  COVID-19 mitigation & therecount, GovRonDeSantis, tatereeves, RandPaul, NYGovCuomo\\
    42 &  State-level policies & MZHemingway, MrAndyNgo, NYPDnews, CurtisHouck, NYCMayorsOffice\\
    46 &  Trump impeachment & therecount, RandPaul, cspan, JackPosobiec, TrumpWarRoom\\
     49 &  Mixed family-related news & FLOTUS, mindykaling, CaslerNoel, tomselliott, DavidLimbaugh\\
     54 & Cuomo assault allegations & LindseyBoylan, BreitbartNews, AOC, rosemcgowan, DailyCaller\\
    55 &  Celebrities & LilNasX, tedcruz, BillyBaldwin, justinbieber, goldenglobes\\
    58 &  Sinophobic crimes & CNN, NRA, BarackObama, TimJGraham, TODAYshow\\
    65 & Congress votes & mtgreenee, GOPLeader, AOC, Liz\_Cheney, JakeSherman\\
     72 & Travel, Covid-19 related & FAANews, alaskalandmine, AmericanAir, thehill, united\\
    73 & Boxing/MMA & espnmma, MarkPrinceOBE, MMAjunkie, mmadissect, UFCEurope\\
    \bottomrule
\end{tabular}
\caption{Top 5 most frequent Twitter accounts embedded in articles per topic. We only show topics that were found to have significantly more embedded tweets on average. Note, this includes both articles from mainstream and fringe outlets. Due to fringe outlets publishing many more articles with embedded tweets, the most frequent accounts embedded are biased towards the actors embedded in fringe articles.}
\label{tab:topaccounts_topics}
\end{table*}



\section{Discussion}
In this paper, we examined the practice of embedding tweets in online news articles. In particular, we compared this practice across reliable news outlets and fringe news outlets. We found that fringe news outlets embedded tweets more frequently than reliable outlets, and that fringe outlets frequently use embedded tweets as the source of the article rather than as illustrations in the article.

Overall, the general types of tweets, accounts, and topics embedded in online news articles were not extremely different across outlet types. While we see certain functions of tweets used significantly more or less by each outlet type (see Table \ref{tab:codes_outlettype}), all functions were represented in each outlet group. For example, when both reliable and fringe outlets embedded tweets, it was most often an elite opinion on an issue, particularly when the article topic was political. Given what we know from the literature, these results should be expected.  \citet{oschatz2021twitter} demonstrated that elite actors are over-represented in the use of embedded tweets in news articles, suggesting that the power asymmetries that existed in the past still exist in today's hybrid media systems \citep{langer2021political}. Our findings extend this result to fringe media, who embedded elite opinions even more so than reliable media. These findings also relate to indexing theory, which suggests that journalists will report on elite perspectives \cite{bennett1990toward, muddiman2022indexing}. Although further work is needed to understand when and how often embedded social media reflects sharing elite perspectives (indexing theory) versus challenging elite perspectives (watchdog journalism). 

We also found that both types of outlets use embedded tweets to syndicate and cite other news items from others or themselves. News syndication and imitation has long been practiced in journalism \cite{breed1955newspaper, boczkowski2010news, starbird2018ecosystem}, and prior work has shown that fringe news outlets often copy articles from mainstream outlets \citep{horne2019different}. However, this practice is typically done through quoting from, copying from, or citing news articles, not tweets. This result is made more interesting when digging into the tweets in our `Tweet Syndication' and `Internal Link' categories. These tweets are not always breaking news from another news organization (as is typical in syndication), but are sometimes tweets with embedded videos from cable news coverage or further links to other news articles. This result clearly demonstrates just how hybrid our media systems are. Further, it is another example of the normalization of social media to fit prior professional norms and practices by journalists \cite{lasorsa2012normalizing}.

Lastly, we did find some unique practices by fringe outlets that merit further study. Most notably, we found that tweets are often used in ``curated lists'' consisting of multiple opinions on a topic. In these cases, multiple tweets are selected by the writers and framed as either a debate between elites or a consensus between elites, yet, the accounts themselves may never have interacted on Twitter and the context around a discussion can be stripped away. In this way, tweets can be cherry-picked for support of specific narratives.

There are some clear dangers with this practice. As put by \citet{oschatz2021twitter}: ``A single tweet can hardly represent a broader discussion in the Twitterverse (e.g., \#metoo, \#coronavirus). It is simply a snapshot of a more or less elaborated opinion of one individual and might skew public opinion.'' We would further argue that with fringe media, there is not only a danger that one type of opinion is over-represented, but that there is a danger that a single tweet can be put into a new context and remixed to fuel harmful narratives \cite{gruppi2021tweeting}. The literature on media manipulation has expressed this concern more generally. For example, \citet{krafft2020disinformation} note that ``[d]ecentralized communication infrastructures, like the web and platforms, foster decontextualization by allowing networks to fork conversations in different ways, creating fragmented publics.'' The use of tweets in news articles is just one instance of this larger issue of decontextualization in hybrid media. We saw some examples of this practice in our data, but we think further work is needed to fully flesh out this tactic. For example, understanding what types of events or narratives are more likely to decontextualize tweets would require a different inductive coding scheme and strategy to confidently assess.

\subsection{Alternative Explanations for Findings}
Due to our data sampling, there are some important alternative explanations for our findings that should be considered. First, due to the random sampling within the reliable and unreliable labeled outlet groups, our sample has ideological bias, where the ``reliable'' group contains mostly politically-left and centrist news outlets and the ``fringe'' group contains mostly politically-right and far-right news outlets. Hence, some of the differences in the use and function of tweets could be due to the difference between political-leaning, rather than veracity. Further, these labels are making broad assumptions of veracity across media outlets, which may vary article by article. This bias points to the larger issue of using weak-labels, such as those from third-party sites like MBFC, to study online news. This concern has most prominently been pointed out in the literature on automated news veracity detection \cite{bozarth2020higher}.

Second, these differences could be explained by the origins of fringe outlets; namely, fringe sources are digitally born. All 9 outlets that we examined in the fringe category were founded as websites, the earliest two being founded in 1998 and 2004, only a few years before the creation of Twitter. While several of the fringe outlets have cable components now, none of them started with television news broadcasts, and none have ever had a traditional print component to their operations. On the other hand, all 9 of the reliable outlets examined in this study began online news production late into their existence, starting out as exclusively operating in print, radio, or cable mediums. Furthermore, many of the writers for the fringe outlets have no formal journalistic training and started out as citizen journalist or bloggers before being hired by these organizations. Hence, the fringe outlets and those who write for the fringe outlets are, for lack of a better term, online digital media natives. Hence, the use of embedded tweets in online news articles could be dependent on if the outlet is legacy media or digital native media. This explanation aligns with the previous results from work by \citet{bane2019tweeting}. Additionally, the lack of traditional training and values may also explain these differences, as sourcing articles from tweets is an inexpensive, low-effort way to produce content. 


\section{Limitations}
This work is not without limitations. First, our qualitative coding used a relatively small sample of the extracted dataset. While these were randomly selected, they were filtered by the Structured Topic Model. This process may have filtered out rare uses and functions of embedded tweets that appeared in topics that used significantly less tweets. However, given the results from our large scale analysis align well with our small scale qualitative analysis, we are confident in these results. 


Second, and perhaps most importantly, Twitter has changed significantly during and after the data used in this study, which may impact how Twitter is used in journalism in the future. During the time of our dataset, Twitter expanded content moderation rules and deplatformed several key right-wing political accounts, like Donald Trump. Notably, despite Trump being banned from Twitter on January 8th, 2021, his tweets were still embedded in news articles well after the fact. However, it is likely that his tweets were covered much more frequently prior to this time frame. Since this study, Twitter was bought by Elon Musk and continues to go through drastic changes under Musk's leadership. At the time of writing, these changes included rule changes around verification on Twitter and proposed changes in content moderation. These changes may impact the platform's future use by journalist. In future work, it will be important understand the impact of content moderation changes, both before and after Musk's takeover, on the practice of embedding tweets in news articles. Further, depending on Twitter's future, another social media platform may takeover the role that Twitter currently plays in journalism.

\section{Broader Perspectives and Ethics Statement}
\paragraph{Potential Negative Outcomes}
Overall, the potential negative outcomes from this study are minimal. However, more generally, there are some concerns that by studying fringe media we may provide them with more attention or we may expose methods that fringe outlets can use to appear more reliable, credible, or mainstream. Our hope is that the benefits of understanding fringe media practices will outweigh this potential harm.

\paragraph{Exposed Accounts}
In this paper, we considered anonymizing the Twitter accounts embedded in news articles (e.x. Figure \ref{fig:topaccounts}). However, by doing so we would destroy the provenance of our findings. More importantly, there were no ``general public'' accounts or private accounts in the top 25 embedded accounts, only verified public figures or organizations. While we did see some public opinion represented by embedded tweets in our study (as described in Table \ref{tab:codes}), we do not expose those accounts in any figure, table, or discussion to preserve privacy. 

\section{Conclusion}
As Twitter, and social media more broadly, continues to be used in news coverage, understanding its role is salient. This study adds to the evidence that elite actors are over-represented in hybrid media, potentially reducing the representation of grass-roots movements and civic dialog in news coverage. Compared to previous work, we dissect the functions of embedded social media into granular categories, providing new insights into previously overlooked practices. Methodologically, we hope that our use of both large-scale, quantitative analysis and small-scale, qualitative analysis provides support for the use of mixed-methods in future studies of journalism practices. 





\begin{small}
\bibliography{references}

\begin{thebibliography}{46}
\providecommand{\natexlab}[1]{#1}

\bibitem[{Althoff, Jindal, and Leskovec(2017)}]{althoff2017online}
Althoff, T.; Jindal, P.; and Leskovec, J. 2017.
\newblock Online actions with offline impact: How online social networks
  influence online and offline user behavior.
\newblock In \emph{Proceedings of the tenth ACM international conference on web
  search and data mining}, 537--546.

\bibitem[{Bane(2019)}]{bane2019tweeting}
Bane, K.~C. 2019.
\newblock Tweeting the agenda: How print and alternative web-only news
  organizations use Twitter as a source.
\newblock \emph{Journalism Practice}, 13(2): 191--205.

\bibitem[{Bennett(1990)}]{bennett1990toward}
Bennett, W.~L. 1990.
\newblock Toward a theory of press-state.
\newblock \emph{Journal of communication}, 40(2): 103--27.

\bibitem[{Blach-{\O}rsten, Eberholst, and Burkal(2017)}]{blach2017hybrid}
Blach-{\O}rsten, M.; Eberholst, M.~K.; and Burkal, R. 2017.
\newblock From hybrid media system to hybrid-media politicians: Danish
  politicians and their cross-media presence in the 2015 national election
  campaign.
\newblock \emph{Journal of Information Technology \& Politics}, 14(4):
  334--347.

\bibitem[{Blei, Ng, and Jordan(2003)}]{blei2003latent}
Blei, D.~M.; Ng, A.~Y.; and Jordan, M.~I. 2003.
\newblock Latent dirichlet allocation.
\newblock \emph{Journal of machine Learning research}, 3(Jan): 993--1022.

\bibitem[{Boczkowski(2010)}]{boczkowski2010news}
Boczkowski, P.~J. 2010.
\newblock \emph{News at work: Imitation in an age of information abundance}.
\newblock University of Chicago Press.

\bibitem[{Bozarth, Saraf, and Budak(2020)}]{bozarth2020higher}
Bozarth, L.; Saraf, A.; and Budak, C. 2020.
\newblock Higher ground? How groundtruth labeling impacts our understanding of
  fake news about the 2016 US presidential nominees.
\newblock In \emph{Proceedings of the International AAAI Conference on Web and
  Social Media}, volume~14, 48--59.

\bibitem[{Breed(1955)}]{breed1955newspaper}
Breed, W. 1955.
\newblock Newspaper ‘opinion leaders’ and processes of standardization.
\newblock \emph{Journalism Quarterly}, 32(3): 277--328.

\bibitem[{Broersma and Graham(2012)}]{broersma2012social}
Broersma, M.; and Graham, T. 2012.
\newblock Social media as beat: Tweets as a news source during the 2010 British
  and Dutch elections.
\newblock \emph{journalism practice}, 6(3): 403--419.

\bibitem[{Broersma and Graham(2013)}]{broersma2013twitter}
Broersma, M.; and Graham, T. 2013.
\newblock Twitter as a news source: How Dutch and British newspapers used
  tweets in their news coverage, 2007--2011.
\newblock \emph{Journalism practice}, 7(4): 446--464.

\bibitem[{Chadwick(2017)}]{chadwick2017hybrid}
Chadwick, A. 2017.
\newblock \emph{The hybrid media system: Politics and power}.
\newblock Oxford University Press.

\bibitem[{Chan(2017)}]{chan2017media}
Chan, M. 2017.
\newblock Media use and the social identity model of collective action:
  Examining the roles of online alternative news and social media news.
\newblock \emph{Journalism \& Mass Communication Quarterly}, 94(3): 663--681.

\bibitem[{Dumitrescu and Ross(2021)}]{dumitrescu2021embedding}
Dumitrescu, D.; and Ross, A.~R. 2021.
\newblock Embedding, quoting, or paraphrasing? Investigating the effects of
  political leaders’ tweets in online news articles: The case of Donald
  Trump.
\newblock \emph{new media \& society}, 23(8): 2279--2302.

\bibitem[{Farhall et~al.(2019)Farhall, Carson, Wright, Gibbons, and
  Lukamto}]{farhall2019political}
Farhall, K.; Carson, A.; Wright, S.; Gibbons, A.; and Lukamto, W. 2019.
\newblock Political elites’ use of fake news discourse across communications
  platforms.
\newblock \emph{International Journal of Communication}, 13.

\bibitem[{Farrar-Myers and Vaughn(2015)}]{farrar2015controlling}
Farrar-Myers, V.~A.; and Vaughn, J.~S. 2015.
\newblock \emph{Controlling the message: New media in American political
  campaigns}.
\newblock NYU Press.

\bibitem[{Flaxman, Goel, and Rao(2016)}]{flaxman2016filter}
Flaxman, S.; Goel, S.; and Rao, J.~M. 2016.
\newblock Filter bubbles, echo chambers, and online news consumption.
\newblock \emph{Public opinion quarterly}, 80(S1): 298--320.

\bibitem[{Freelon, McIlwain, and Clark(2018)}]{freelon2018quantifying}
Freelon, D.; McIlwain, C.; and Clark, M. 2018.
\newblock Quantifying the power and consequences of social media protest.
\newblock \emph{New Media \& Society}, 20(3): 990--1011.

\bibitem[{Fuchs(2010)}]{fuchs2010alternative}
Fuchs, C. 2010.
\newblock Alternative media as critical media.
\newblock \emph{European journal of social theory}, 13(2): 173--192.

\bibitem[{Gruppi et~al.(2021)Gruppi, Adal{\i}, Salemi, and
  Horne}]{gruppi2021tweeting}
Gruppi, M.; Adal{\i}, S.; Salemi, M.; and Horne, B.~D. 2021.
\newblock From Tweeting About News to Creating News Around Tweets:
  Characterizing Tweets Embedded in News Articles.

\bibitem[{Gruppi, Horne, and Adal{\i}(2022)}]{gruppi2022nela}
Gruppi, M.; Horne, B.~D.; and Adal{\i}, S. 2022.
\newblock NELA-GT-2021: A Large Multi-Labelled News Dataset for The Study of
  Misinformation in News Articles.
\newblock \emph{arXiv preprint arXiv:2203.05659}.

\bibitem[{Guess et~al.(2018)Guess, Nyhan, Lyons, and
  Reifler}]{guess2018avoiding}
Guess, A.; Nyhan, B.; Lyons, B.; and Reifler, J. 2018.
\newblock Avoiding the echo chamber about echo chambers.
\newblock \emph{Knight Foundation}, 2: 1--25.

\bibitem[{Harcup and O’neill(2017)}]{harcup2017news}
Harcup, T.; and O’neill, D. 2017.
\newblock What is news? News values revisited (again).
\newblock \emph{Journalism studies}, 18(12): 1470--1488.

\bibitem[{Hedman(2015)}]{hedman2015j}
Hedman, U. 2015.
\newblock J-Tweeters: Pointing towards a new set of professional practices and
  norms in journalism.
\newblock \emph{Digital Journalism}, 3(2): 279--297.

\bibitem[{Hedman and Djerf-Pierre(2013)}]{hedman2013social}
Hedman, U.; and Djerf-Pierre, M. 2013.
\newblock The social journalist: Embracing the social media life or creating a
  new digital divide?
\newblock \emph{Digital journalism}, 1(3): 368--385.

\bibitem[{Heravi and Harrower(2016)}]{heravi2016twitter}
Heravi, B.~R.; and Harrower, N. 2016.
\newblock Twitter journalism in Ireland: sourcing and trust in the age of
  social media.
\newblock \emph{Information, communication \& society}, 19(9): 1194--1213.

\bibitem[{Holt, Ustad~Figenschou, and Frischlich(2019)}]{holt2019key}
Holt, K.; Ustad~Figenschou, T.; and Frischlich, L. 2019.
\newblock Key dimensions of alternative news media.
\newblock \emph{Digital Journalism}, 7(7): 860--869.

\bibitem[{Horne, N{\o}rregaard, and Adal{\i}(2019)}]{horne2019different}
Horne, B.~D.; N{\o}rregaard, J.; and Adal{\i}, S. 2019.
\newblock Different Spirals of Sameness: A Study of Content Sharing in
  Mainstream and Alternative Media.
\newblock In \emph{Proceedings of the International AAAI Conference on Web and
  Social Media}, volume~13, 257--266.

\bibitem[{Kannan et~al.(2014)Kannan, Gurusamy, Vijayarani, Ilamathi, Nithya,
  Kannan, and Gurusamy}]{kannan2014preprocessing}
Kannan, S.; Gurusamy, V.; Vijayarani, S.; Ilamathi, J.; Nithya, M.; Kannan, S.;
  and Gurusamy, V. 2014.
\newblock Preprocessing techniques for text mining.
\newblock \emph{International Journal of Computer Science \& Communication
  Networks}, 5(1): 7--16.

\bibitem[{Khondker(2011)}]{khondker2011role}
Khondker, H.~H. 2011.
\newblock Role of the new media in the Arab Spring.
\newblock \emph{Globalizations}, 8(5): 675--679.

\bibitem[{KhudaBukhsh et~al.(2021)KhudaBukhsh, Sarkar, Kamlet, and
  Mitchell}]{khudabukhsh2021fringe}
KhudaBukhsh, A.~R.; Sarkar, R.; Kamlet, M.~S.; and Mitchell, T.~M. 2021.
\newblock Fringe news networks: dynamics of US news viewership following the
  2020 presidential election.
\newblock \emph{arXiv preprint arXiv:2101.10112}.

\bibitem[{Krafft and Donovan(2020)}]{krafft2020disinformation}
Krafft, P.; and Donovan, J. 2020.
\newblock Disinformation by design: The use of evidence collages and platform
  filtering in a media manipulation campaign.
\newblock \emph{Political Communication}, 37(2): 194--214.

\bibitem[{Kwak et~al.(2010)Kwak, Lee, Park, and Moon}]{kwak2010twitter}
Kwak, H.; Lee, C.; Park, H.; and Moon, S. 2010.
\newblock What is Twitter, a social network or a news media?
\newblock In \emph{Proceedings of the 19th international conference on World
  wide web}, 591--600.

\bibitem[{Langer and Gruber(2021)}]{langer2021political}
Langer, A.~I.; and Gruber, J.~B. 2021.
\newblock Political agenda setting in the hybrid media system: Why legacy media
  still matter a great deal.
\newblock \emph{The International Journal of Press/Politics}, 26(2): 313--340.

\bibitem[{Lasorsa, Lewis, and Holton(2012)}]{lasorsa2012normalizing}
Lasorsa, D.~L.; Lewis, S.~C.; and Holton, A.~E. 2012.
\newblock Normalizing Twitter: Journalism practice in an emerging communication
  space.
\newblock \emph{Journalism studies}, 13(1): 19--36.

\bibitem[{Lewis, Kaufhold, and Lasorsa(2010)}]{lewis2010thinking}
Lewis, S.~C.; Kaufhold, K.; and Lasorsa, D.~L. 2010.
\newblock Thinking about citizen journalism: The philosophical and practical
  challenges of user-generated content for community newspapers.
\newblock \emph{Journalism practice}, 4(2): 163--179.

\bibitem[{Muddiman et~al.(2022)Muddiman, Budak, Murray, Kim, and
  Stroud}]{muddiman2022indexing}
Muddiman, A.; Budak, C.; Murray, C.; Kim, Y.; and Stroud, N.~J. 2022.
\newblock Indexing theory during an emerging health crisis: how US TV news
  indexed elite perspectives and amplified COVID-19 misinformation.
\newblock \emph{Annals of the International Communication Association}, 46(3):
  174--204.

\bibitem[{Munger and Phillips(2022)}]{munger2022right}
Munger, K.; and Phillips, J. 2022.
\newblock Right-wing YouTube: a supply and demand perspective.
\newblock \emph{The International Journal of Press/Politics}, 27(1): 186--219.

\bibitem[{Olausson(2017)}]{olausson2017reinvented}
Olausson, U. 2017.
\newblock The reinvented journalist: The discursive construction of
  professional identity on Twitter.
\newblock \emph{Digital Journalism}, 5(1): 61--81.

\bibitem[{Oschatz, Stier, and Maier(2021)}]{oschatz2021twitter}
Oschatz, C.; Stier, S.; and Maier, J. 2021.
\newblock Twitter in the News: An Analysis of Embedded Tweets in Political News
  Coverage.
\newblock \emph{Digital Journalism}, 1--20.

\bibitem[{Parmelee(2013)}]{parmelee2013political}
Parmelee, J.~H. 2013.
\newblock Political journalists and Twitter: Influences on norms and practices.
\newblock \emph{Journal of Media Practice}, 14(4): 291--305.

\bibitem[{Roberts et~al.(2014)Roberts, Stewart, Tingley, Lucas, Leder-Luis,
  Gadarian, Albertson, and Rand}]{roberts2014structural}
Roberts, M.~E.; Stewart, B.~M.; Tingley, D.; Lucas, C.; Leder-Luis, J.;
  Gadarian, S.~K.; Albertson, B.; and Rand, D.~G. 2014.
\newblock Structural topic models for open-ended survey responses.
\newblock \emph{American Journal of Political Science}, 58(4): 1064--1082.

\bibitem[{Rodriguez and Storer(2020)}]{rodriguez2020computational}
Rodriguez, M.~Y.; and Storer, H. 2020.
\newblock A computational social science perspective on qualitative data
  exploration: Using topic models for the descriptive analysis of social media
  data.
\newblock \emph{Journal of Technology in Human Services}, 38(1): 54--86.

\bibitem[{Starbird et~al.(2018)Starbird, Arif, Wilson, Van~Koevering, Yefimova,
  and Scarnecchia}]{starbird2018ecosystem}
Starbird, K.; Arif, A.; Wilson, T.; Van~Koevering, K.; Yefimova, K.; and
  Scarnecchia, D. 2018.
\newblock Ecosystem or echo-system? exploring content sharing across
  alternative media domains.
\newblock In \emph{Proceedings of the International AAAI Conference on Web and
  Social Media}, volume~12.

\bibitem[{Wallsten(2015)}]{wallsten2015non}
Wallsten, K. 2015.
\newblock Non-elite Twitter sources rarely cited in coverage.
\newblock \emph{Newspaper Research Journal}, 36(1): 24--41.

\bibitem[{Wilson and Starbird(2020)}]{wilson2020cross}
Wilson, T.; and Starbird, K. 2020.
\newblock Cross-platform disinformation campaigns: lessons learned and next
  steps.
\newblock \emph{Harvard Kennedy School Misinformation Review}, 1(1).

\bibitem[{Zhang et~al.(2018)Zhang, Wells, Wang, and Rohe}]{zhang2018attention}
Zhang, Y.; Wells, C.; Wang, S.; and Rohe, K. 2018.
\newblock Attention and amplification in the hybrid media system: The
  composition and activity of Donald Trump’s Twitter following during the
  2016 presidential election.
\newblock \emph{New Media \& Society}, 20(9): 3161--3182.

\end{thebibliography}
\end{small}

\end{document}